\documentclass{article}
\author{Claire Levaillant}
\title{Multi-image quantum encryption scheme using blocks of bit planes and images}

\newcommand{\la}{\lambda}
\newcommand{\lL}{\lceil log_2\,L\rceil}
\newcommand{\lM}{\lceil log_2\,M\rceil}
\newcommand{\nts}{\negthickspace}
\newcommand{\mc}{\mathcal}
\newcommand{\lb}{\lbrace}
\newcommand{\rb}{\rbrace}
\newcommand{\N}{\mathcal{N}}
\newcommand{\lc}{\lceil}
\newcommand{\rc}{\rceil}

\usepackage{amsmath, amssymb, amsthm, epsfig}

\begin{document}

\maketitle

\section{Introduction, novel multi-image quantum representation and cryptographic scheme outline}
\subsection{Introduction}
With the development of the internet, the security of image information has become an increasingly important issue. Due to the strong correlations among adjacent pixels, bulky data volume and the handling of various data formatting, traditional encryption schemes are not suitable for image encryption. While many protocols exist for single image encryption, the complete shuffling of several images within one another in terms of pixel values is generally not studied, whether classically or in a quantum setting. Because multiple images can be represented as a superposition and unitary transformations simultaneously affect each element of the superposition, the quantum setting is extremely well suited for encrypting multiple images non independently and efficiently using a massive parallel processing. The storage capacity is also exponentially improved. Furthermore, we will see that an interesting feature of encrypting multiple images in a quantum setting is that the attackant has no way of knowing the precise number of images that get encrypted. 
\subsection{Novel multi-image quantum representation}
In \cite{LE}, a new scheme to encode multiple images at once is presented in a quantum setting. A geometric transformation of the square, namely the baker map, which has a quantum implementation in terms of SWAP gates and controlled SWAP gates under some conditions on its parameters \cite{BA} is used to shuffle both the pixel positions and the plane consisting of the images and the bit planes. On one hand, in the case when the number of images to transmit exceeds the number of bit planes, we must add some bit planes that are filled with $0$ bits. On the other hand, if the number of images is less than the number of bit planes (usually $8$), then we must add some blank images. In \cite{ZH}, a scheme for multi-image quantum encryption is provided based on the quantum $3D$ Arnold transform for the geometric scrambling. For an image of size $2^n\times 2^n$ and given $M$ images to encode, the authors must add $2^n-M$ blank images to the quantum representation of the multi-image. In terms of qubits, it means adding $n-\lceil log_2\,M\rceil$ qubits. For instance, when the images are sized $256\times 256$ and there are four images, a total of $6$ qubits needs to be added. Whereas, in \cite{LE} if $L$ denotes the number of bit planes supposedly bigger than the number $M$ of images to be sent, only $\lceil log_2\,L\rceil -\lceil log_2\,M\rceil$ qubits must be added in order to perform the quantum geometric transformation. Taking for instance $L=8$, there will be only one extra qubit in the quantum representation.
The current paper is concerned with transmitting confidentially to the receiver a large set of images at once and optimizing the number of qubits. An idea is to use blocks. The concept of blocks for quantum image encryption originated in \cite{QB}, where the authors divide the plane of pixels into blocks. They use qubits to represent the position of the blocks and the position of the pixels in each block. Here, we divide the set of $M$ images into subsets of $2^{\lceil log_2\,L\rceil}$ images by considering $2^{\lM-\lL}$ blocks, each containing 
$2^{\lL}$ images, where the last blocks are possibly filled or partly filled with blank images. By doing so, we are able to independently scramble the images belonging to a same block and the bit planes. In the quantum representation, an image appears as the position of the block it belongs to and its position inside the block. The total number of qubits needed for a $2^n\times 2^n$ image is thus
$2n+\lL+\lM$ instead of $2n+2\lM$. Thus, if we have $8$ bit planes and $200$ images of size $256\times 256$ to encode, our block representation of quantum multi-image uses a total of $28$ qubits, instead of $33$ qubits in \cite{LE}. \\
We introduce the quantum block bit plane representation for a multi-image, abbreviated QBBRMI, in which a quantum multi-image $|S>$ is represented as:
$$|S>=\la\sum_{x,y=0}^{2^n-1}\sum_{m,l=0}^{2^{\lL}-1}\sum_{t=0}^{2^{\lM-\lL}-1}|P_{tmxyl}>|t>|ml>|xy>$$
with $$\la:=\frac{1}{\sqrt{2^{2n+\lL+\lM}}}.$$
In this representation, $P_{tmxyl}$ is the bit value of the $m$-th image belonging to the $t$-th block at pixel $(x,y)$ on the $l$-th bit plane. This quantum representation importantly allows to retrieve the classical images accurately by projective measurements. 
Based on our present needs, it is inspired from the quantum representation model for multi-image QRMMI of \cite{ZH}, the bit plane representation of quantum images BRQI of \cite{BP}, the bit plane representation for quantum multi-image BRQMI of \cite{LE} and the quantum block image representation QBIR of \cite{QB}. Both QRMMI and QBIR are based on the novel enhanced quantum representation NEQR of \cite{NE}, in which one qubit is used for each bit plane. Instead, the bit plane representation model uses $\lL$ qubits. Moreover, it was introduced by the authors of \cite{BP} to allow for independent bit-plane scrambling of the pixel positions. Our novel model allows for each pixel position to scramble independently the bit planes and the images belonging to a same block on one hand and for a given block, a given image belonging to that block and a given bit plane to scramble the pixel positions. For the scrambling, we use independent quantum baker maps whose partition parameters and iteration parameters depend on the block and the pixel position in the first case and on the block, the image position inside the block and the bit plane in the second case. The advantage of the baker map versus the Arnold transform is that it has a longer scrambling period.
\subsection{Cryptographic scheme outline}
Once the bit values of the images are all scrambled, the pixel correlation is in particular entirely broken. We further diffuse the images using the sine chaotification of a 5D hyper chaotic map. Hyperchaos was first reported by R\"ossler. Hyperchaotic systems have more than one positive Lyapunov exponent and have more complicated dynamics than ordinary chaotic ones. Chaotic maps are used for diffusion in image encryption as they are able to generate completely different sequences through relatively small changes in the tuning parameters and in the initial conditions. The idea to use them for cryptographic applications originates in the paper by Matthews \cite{MA}. Since then, many image cryptographic protocol have been based on the use of chaotic maps and sometimes their combinations. Following this idea of composing several chaotic maps, we apply here a sine chaotification to the 5D hyper chaotic map. The composition of these two maps increases the security of the scheme as far as key sensitivity and allowing the key space to be infinite. The sine chaotification model was originally introduced in \cite{HZ} and presents three major advantages when combined with another chaotic map. Firstly, it enlarges the Lyapunov exponent of the system, hence makes the system even more disordered and the sensitivity to the initial conditions gets reinforced. The key sensitivity is then excellent. This is crucial as a decryption using a very similar key should completely fail to recover the plaintext image. Secondly and additionally, the enlargement of the Lyapunov exponent may be controlled. Third, the range of the tuning parameters for chaos becomes infinite. For cryptographic applications, it means an infinite key space and schemes thus invulnerable to brute force attacks. 
Chaotic map are traditionally used in image encryption due to sensitivity to initial conditions to produce pseudo-random bit sequences. In our scheme, the initial conditions depend on the plaintext images so as to avoid chosen plaintext attacks and chosen ciphertext attacks. A slight bit value change in the total set of images will completely modify the ciphertext multi-image. The pseudo-random sequences issued from the sine chaotification of the 5D hyperchaotic system get used together with some additional chaotic Chebyshev polynomial map to produce secret bit values which get XORed with the bit values of the images. In \cite{LE}, the images get diffused independently by setting different control parameters for each image in the chaotic map. This is alright in the case when the number of images to encrypt is relatively small. As the current paper rather addresses a larger set of images to be transmitted, this way would mean too many additional secret keys. Thus, the tuning parameters of the system get set independently from the scrambled images. This is permitted by the higher dimension of the chaotic system. This higher dimension of the chaotic system allows for an overall diffusion of the scrambled images, instead of having each scrambled image diffused independently. \\
In our scheme, the security level is enhanced versus other schemes by using a higher dimensional hyper chaotic map, by coupling two chaotic systems, namely the sine chaotification model and the 5D hyper chaotic map, and by cascading another chaotic map, namely the Chebyshev transformation. 

\subsection{Paper outline}

Our paper is organized as follows. In the next part, we provide a complete description of the cryptographic scheme. We introduce the quantum baker map and show how to use it for the scrambling of the images. We then deal with the diffusion process by first introducing the sine chaotification of a 5D hyperchaotic system and then explaining how to produce the secret bit values. The last part of the paper addresses the quantum computational complexity, that is the complexity of the quantum circuit used during the quantum encryption and the quantum decryption.  As part of our results, we exhibit a general quantum circuit for the quantum baker map, depending on the values of its parameters. We thus obtain a precise count for the number of quantum gates needed during the scrambling process. 
It allows us to draw comparisons between the scheme of \cite{LE} and the one presented here in terms of quantum circuit depth and width. We finally conclude the paper by pointing out that the new scheme does not rule out the old one. Both schemes should rather be used depending on various specificities of the plaintext which we address and the precise protocol we use. 

\section{Encryption and decryption of the multi-image}

\subsection{Scrambling with the quantum baker map}
We first introduce briefly the quantum baker map used for scrambling. The definition and theorem below are both taken from \cite{BA}. 

\newtheorem{Theorem}{Theorem}
\newtheorem{Definition}{Definition}

\begin{Definition} The discrete baker map is defined over a square of size $2^n=2^{q_1}+\dots + 2^{q_k}$ by:
$$\left\lbrace\begin{array}{l}
N_0=0\; \text{and}\; N_i=2^{q_1}+\dots +2^{q_i}\;\text{and}\;
x=N_{i-1},N_{i-1}+1,\dots,N_i-1\\
(x^{'},y^{'})=\Bigg(2^{n-q_i}(x-N_{i-1})+y\,\text{mod}\,2^{n-q_i},N_{i-1}+\frac{y-y\,\text{mod}\,2^{n-q_i}}{2^{n-q_i}}\Bigg)
\end{array}\right.$$
\end{Definition}

\begin{Theorem} Quantum implementation of the discrete baker map. \\
(i) The map defined by
$$M_s(x,y)=\Bigg( 2^{n-s}x\,\text{mod}\,2^n+y\,\text{mod}\,2^{n-s},\frac{y-y\,\text{mod}\,2^{n-s}}{2^{n-s}}+x-x\,\text{mod}\,2^s\Bigg)$$
has a quantum implementation. \\
(ii) Each subfunction on $\lbrace N_{i-1},\dots,N_i-1\rbrace$ of the discrete baker map is $M_{q_i}$ if and only if $2^{q_i}|2^{q_1}+\dots+2^{q_{i-1}}$.
\end{Theorem}

In what follows, we will only consider discrete baker maps with parameters $q_1,\dots, q_k$ satisfying to the conditions $2^{q_i}|2^{q_1}+\dots +2^{q_{i-1}}$ for each $i$ with $2\leq i\leq k$. We call such discrete baker maps ``quantum baker map" since by Theorem $1$ due to Hou and Liu they have a quantum implementation. \\
Like in \cite{LE}, we operate a two-stage scrambling, each time using quantum baker maps. Our scrambling is block independent, namely for each block, the parameters of the quantum baker map and the number of iterations are picked differently. More explicitly, for each pixel $(x,y)$ we do a block independent scrambling of the images and of the bit planes. We obtain:

\begin{eqnarray*}|S^{'}>&\,\nts\nts\nts\nts=&\nts\nts\la\sum_{x,y=0}^{2^n-1}\sum_{m,l=0}^{2^{\lL}-1}\sum_{t=0}^{2^{\lM-\lL}-1}\nts\nts\nts\nts\nts\nts\nts\nts|P_{tmxyl}>|t>|xy>QBM_{x,y,t}^{r(x,y,t)}(|ml>)\\
&\,\nts\nts\nts\nts=&\nts\nts\la\sum_{x,y=0}^{2^n-1}\sum_{m,l=0}^{2^{\lL}-1}\sum_{t=0}^{2^{\lM-\lL}-1}\nts\nts\nts\nts\nts\nts\nts\nts|P^{'}_{tmxyl}>|t>|xy>|ml>\end{eqnarray*}
We then fix a bit plane and an image belonging to a certain block and perform independent scramblings of the pixel positions. It yields:
\begin{eqnarray*}|S^{''}>&\,\nts\nts\nts\nts=&\nts\nts\la\sum_{x,y=0}^{2^n-1}\sum_{m,l=0}^{2^{\lL}-1}\sum_{t=0}^{2^{\lM-\lL}-1}\nts\nts\nts\nts\nts\nts\nts\nts\nts|P_{tmxyl}>|t>|ml>QBM_{l,m,t}^{r(l,m,t)}(|xy>)\\
&\,\nts\nts\nts\nts=&\nts\nts\la\sum_{x,y=0}^{2^n-1}\sum_{m,l=0}^{2^{\lL}-1}\sum_{t=0}^{2^{\lM-\lL}-1}\nts\nts\nts\nts\nts\nts\nts\nts\nts|P^{''}_{tmxyl}>|t>|xy>|ml>\end{eqnarray*}
From there, the scrambled quantum multi-image is ready to be diffused. 
\subsection{Diffusion using a five dimensional hyperchaotic map}
According to \cite{SS}, the 5-D hyperchaotic system
$$\left\lbrace\begin{array}{l}
\overset{.}{x_1}=a(x_2-x_1)+x_2x_3x_4\\
\overset{.}{x_2}=b(x_1+x_2)+x_5-x_1x_3x_4\\
\overset{.}{x_3}=-cx_2-dx_3-ex_4+x_1x_2x_4\\
\overset{.}{x_4}=-fx_4+x_1x_2x_3\\
\overset{.}{x_5}=-g(x_1+x_2)
\end{array}\right.$$
when the control parameters are set to $a=30,b=10,c=15.7,d=5,e=2.5,f=4.45$ and $g=38.5$ is in a chaotic state and could produce five chaotic sequences. Some phase portraits are displayed in Fig. $1$ of \cite{SS}. The randomness performance of a chaotic system gets measured through the Lyapunov exponent. A positive Lyapunov exponent means the chaotic behavior and a larger Lyapunov corresponds to higher sensitivity to the initial conditions. An idea to increase the Lyapunov exponents and also be able to make them as large as desired is to use a sine chaotification model. The idea originated in \cite{HZ}. Namely, given $\phi$ a chaotic map, create a chaotic system 
$$x_{n+1}=h(x_n)=sin(\pi\lambda\phi(x_n)),$$
where $\lambda$ is any parameter such that $\lambda\geq 1$. As a matter of fact, 
$$LE_h=LE_g+LE_{\phi}+ln(\lambda)>LE_{\phi},$$
where $g(x)=sin(\pi x)$, see \cite{HZ}. 
Another advantage of the sine chaotification model is to make the control parameter range infinite and thus for our cryptographic applications make the key space infinite. 
Based on these considerations, we define a 5-D chaotic system:
$$\left\lbrace\begin{array}{l}
\overset{.}{x_1}=sin(\pi\la_1(30(x_2-x_1)+x_2x_3x_4))\\
\overset{.}{x_2}=sin(\pi\la_2(10(x_1+x_2)+x_5-x_1x_3x_4)))\\
\overset{.}{x_3}=sin(\pi\la_3(-15.7x_2-5x_3-2.5x_4+x_1x_2x_4))\\
\overset{.}{x_4}=sin(\pi\la_4(-4.45x_4+x_1x_2x_3))\\
\overset{.}{x_5}=sin(\pi\la_5(-38.5(x_1+x_2)))
\end{array}\right.$$

Our seed will be designed to depend on the plaintext images. A slight bit change will completely modify the ciphertext multi-image. We choose $x(0)$ to be the total intensity of the images and then define $z(0):=Mod[x(0)\times 10^3,1]$. These two numbers belong to the interval $[0,1]$. The initial values $y(0)$ and $t(0)$ are then obtained by applying some Chebyshev transformation to respectively $x(0)$ and $z(0)$. We recall that the $k$-th Chebyshev polynomial $T_k$ is the unique polynomial of $\mathbb{R}[X]$ such that:
$$\forall \theta\in\mathbb{R},\,T_k(cos\,\theta)=cos(k\theta)$$
Explicitly, 
$$x(0):=\sum_{\begin{array}{l}0\leq x,y\leq 2^n-1\\0\leq m\leq M\end{array}}\frac{C_{xym}}{M2^{2n}(2^{L}-1)}$$
with $C_{xym}$ the pixel value of the image $m$ at pixel $(x,y)$. 
As for $y(0)$ and $t(0)$, define for instance two plaintext dependent integers:
$$\alpha:=\frac{1}{2^{2n}}\sum_{0\leq x,y\leq 2^n}\sum_{\begin{array}{l}\qquad 0\leq m,l\leq 2^{\lL}-1\\0\leq b\leq 2^{\lM-\lL}-1\end{array}}P_{bmxyl}$$
and $$\beta:=\frac{1}{2^{2n}}\sum_{0\leq x,y\leq 2^n}\left(\begin{array}{l}\sum_{\begin{array}{l}\qquad 0\leq m,l\leq 2^{\lL}-1\\0\leq b\leq 2^{\lM-\lL}-1\end{array}}P_{bmxyl}\end{array}\right)^2$$
That is for a pixel $(x,y)$, consider the total number of bit values equal to $1$ for all the images. Then $\alpha$ is the average of these numbers and $\beta$ is the average of these 
squared numbers.  
Now set:
 \begin{eqnarray*}
y(0)&=&T_{\alpha}(x(0))\\
t(0)&=&T_{\beta}(z(0))
\end{eqnarray*}
Note that the initial conditions want to smartly carry the fact the attackant has no way of knowing how many images are encrypted. 
From there, iterate the dynamical system and ignore the first hundred iterations so as to avoid transient effects. Do so as long as is necessary to get four sequences of distinct numbers:
$$\begin{array}{l}
x_0,\dots,x_{2^n-1}\\
y_0,\dots,y_{2^n-1}\\
z_0,\dots,z_{2^{\lL}-1}\\
t_0,\dots,t_{2^{\lM-\lL}-1}
\end{array}$$

Then sort the four lists by increasing order and assign to the $i$-th (resp $j$-th) pixel coordinate (resp $m$-th image, resp $b$-th block) the index position of $x_i$ (resp $y_j$, resp $z_m$, resp $t_b$) in the sorted sequences. Obtain in turn four sequences of integers:
$$\begin{array}{l}
n_0,\dots,n_{2^n-1}\\
k_0,\dots,k_{2^n-1}\\
r_0,\dots,r_{2^{\lL}-1}\\
s_0,\dots,s_{2^{\lM-\lL}-1}
\end{array}$$

For an image $m$ belonging to a block $b$ and a pixel $(i,j)$ on that image, compute 
\begin{equation*}\begin{split}\lfloor T_{n_i}(y_{2^n-i+1})T_{k_j}(x_{2^n-j+1})T_{r_m}(t_{2^{\lM-\lL}-b+1})T_{s_b}(z_{2^{\lL}-m+1})&\times 10^{10}\rfloor\\&\nts\nts\nts\nts\nts\nts\nts\nts\,\text{mod}\;2^{2^{\lL}}\end{split}\end{equation*}
Obtain a  $2^{\lceil log_2L\rceil}$-qubit secret key
$$|K_{2^{\lL}-1,m,b,i,j}\dots K_{0,m,b,i,j}>.$$
Terminate the encryption process by diffusing independently all the images using qubit ancillas encoding the secret keys in order to perform 
$2^{2n+\lM+\lL}$ CCNOT gates. Obtain the quantum ciphertext:
$$\la\sum_{i,j=0}^{2^n-1}\sum_{m,l=0}^{2^{\lL}-1}\sum_{b=0}^{2^{\lM-\lL}-1}\nts\nts\nts\nts\nts\nts\nts\nts\nts|P^{''}_{blmij}\oplus K_{l,m,b,i,j}>|b>|ij>|ml>$$
This quantum ciphertext is then converted to its classical version using projective measurements and sent to the receiver. In turn, the receiver converts it back into QBBRMI representation and uses his or her quantum computer to perform all the quantum gates in the reverse order of the quantum encryption process. A new set of projective measurements is finally performed in order to retrieve the transmitted classical images. Finally, the blank images if any, get discarded. 

\section{Complexity analysis of the quantum circuit}

In this part, we exhibit a quantum circuit for a general quantum baker map and proceed to a gate count depending on the values of the parameters of the quantum baker map. We then analyse on some examples the number of qubits needed and the depth of the quantum circuit for the scheme of \cite{LE} and for the one presented here.
\subsection{Quantum circuit for the quantum baker map}
In what follows, we consider a quantum baker map with parameters $(q_1,\dots,q_k)$ with $2^{q_i}$ divides $2^{q_1}+\dots + 2^{q_{i-1}}$ for each $i$ with $2\leq i\leq k$ and $2^{q_1}+\dots + 2^{q_k}=2^n$. Let $f$ be the subfunction defined on $[\!|0,\dots,2^{q_1}-1|\!]$. And so, by Theorem $1$, we have $f=M_{q_1}$. Explicitly,
$$f(x,y)=(x_{q_1-1}\dots x_0\,y_{n-q_1-1}\dots y_0,x_{n-1}\dots x_{q_1}\,y_{n-1}\dots y_{n-q_1})$$
Below, we define a few quantum gates which will be used to implement $f$.

\begin{Definition}
\begin{eqnarray*}
A_0&:=&\prod_{i=0}^{n-q_1-1}Sw(x_i,y_i)\\
B_0&:=&\prod_{i=0}^{q_1-1}Sw(x_{n-q_1+i},y_i)\\
B_1&:=&\prod_{i=0}^{q_1-1}Sw(y_{n-q_1+i},y_i)\\
C_0&:=&\prod_{j=0}^{q_1-1}\prod_{i=0}^{n-q_1-1}Sw(y_{n-q_1-i+j},y_{n-q_1-i-1+j})\\
C_1&:=&\prod_{i=q_1}^{n-1}Sw(x_i,y_i)\\
C_2&:=&\prod_{j=0}^{n-q_1-1}\prod_{i=0}^{2q_1-n-1}Sw(x_{q_1-i+j},x_{q_1-i-1+j})
\end{eqnarray*}
\end{Definition}

We are now ready to state the following result.

\begin{Theorem} Quantum implementation of the first subfunction of the quantum baker map.
$$f=\begin{cases}
A_0B_0B_1&\text{if $q_1\leq\frac{n}{2}$}\\
&\\
A_0C_0C_1C_2&\text{if $q_1>\frac{n}{2}$}
\end{cases}$$
\end{Theorem}

The quantum baker map is defined by subfunctions as follows:
$$\begin{array}{l}\text{On}\;[\!\hspace{-0.015cm}|0,2^{q_1}-1|\!\hspace{-0.015cm}]\;\text{the QBM is}\; f_1=f\\
\text{On}\;\Big[\!\hspace{-0.05cm}\Big|\sum_{i=1}^{r-1}2^{q_i},\Big(\sum_{i=1}^{r-1}2^{q_i}\Big)+2^{q_r}-1\Big|\!\hspace{-0.05cm}\Big]\;\text{the QBM is}\;f_r=M_{q_r}\end{array}$$
The quantum circuit for the quantum baker map consists of applying the subfunctions $f_r$ sequentially, with $1\leq r\leq k$.
Like seen above, the quantum circuit for $f_1$ is composed of SWAP gates. The quantum circuit for the $f_r$'s with $r>1$ will rather be composed of controlled SWAP gates. In order to identify the qubits carrying the control, it will be necessary to reduce
$$\sum_{i=1}^{r-1}2^{q_i}=\sum_{s=1}^{l}2^{j_s},$$
so that the new exponents are all distinct integers. Since $$2^{q_r}\Big|\sum_{s=1}^l 2^{j_s},$$ it follows that 
$$q_r\leq Min_{1\leq s\leq l}(j_s).$$Then,
$$x\in\Bigg[\!\hspace{-0.05cm}\Bigg|\sum_{s=1}^{l}2^{j_s},\sum_{s=1}^{l}2^{j_s}+2^{q_r}-1\Bigg|\!\hspace{-0.05cm}\Bigg]\Leftrightarrow\left\lb\begin{array}{l}\forall s\in[\!\hspace{-0.05cm}| 1,l|\!\hspace{-0.05cm}],\;x_{j_s}=1\\\\\forall j\geq q_r\;\text{such that}\; j\not\in\cup_{s=1}^l\lb j_s\rb,\;x_j=0.\end{array}\right.$$

Thus, the controls on the SWAP gates of $f_r$, $r\geq 2$, are defined like follows.
\begin{Definition}\hfill\\
(i) Suppose $2\leq r\leq k-1$, then
\begin{equation*}\begin{split}
C_r:=&\prod_{s=1}^l C_{y_{j_s}=1}^{\lbrace j_s\geq q_1\rbrace}\,C^{\lbrace j_s<q_1\rbrace}_{x_{n-q_1+j_s}=1}\\
&\prod_{\begin{array}{l}\;\;\;\;\;\,j=q_r\\j\not\in\cup_{s=1}^l\lbrace j_s\rbrace\end{array}}^{n-1}C_{y_j=0}^{\lbrace j\geq q_1\rbrace}C_{x_{n-q_1+j}=0}^{\lbrace j<q_1\rbrace}\end{split}\end{equation*}
(ii) \begin{equation*}C_k:=\prod_{s=1}^l C_{y_{j_s}=1}^{\lbrace j_s\geq q_1\rbrace}C_{x_{n-q_1+j_s}=1}^{\lbrace j_s<q_1\rbrace}\end{equation*}
\end{Definition}
Define now the following controlled SWAP gates.
\begin{Definition}
\begin{eqnarray*}
D_0^{(i)}&:=&\prod_{l=0}^{q_i-1}C_iSw(y_l,y_{q_1-q_i+l})\\
D_1^{(i)}&:=&\prod_{l=1}^{q_1-q_i}C_iSw(x_{n-l},y_{q_1-l})\\
D_2^{(i)}&:=&\prod_{l=1}^{q_i}C_iSw(x_{n-l},x_{n-l+q_i-q_1})\\
D_3^{(i)}&:=&\prod_{l=0}^{q_i-1}C_iSw(y_l,x_{n-q_i+l})\\
E_0^{(i)}&:=&\prod_{l=q_i}^{q_1-1}C_iSw(y_l,x_{n-q_1+l})\\
E_1^{(i)}&:=&\prod_{k=0}^{q_1-q_i-1}\prod_{l=0}^{q_i-1}C_iSw(x_{n-q_1+q_i-l+k},x_{n-q_1+q_i-l+k-1})\\
E_2^{(i)}&:=&\prod_{l=0}^{q_1-q_i-1}C_iSw(y_l,x_{n-q_1+l})\\
E_3^{(i)}&:=&\prod_{k=0}^{2q_i-q_1-1}\prod_{l=0}^{q_1-q_i-1}C_iSw(y_{q_1-q_i-l+k},y_{q_1-q_i-l+k-1})\\
F_1^{(i)}&:=&\prod_{k=0}^{q_1-1}\prod_{l=0}^{q_i-q_1-1}C_iSw(x_{n-q_1-l+k},x_{n-q_1-l+k-1})\\
F_2^{(i)}&:=&\prod_{l=0}^{q_i-q_1-1}C_iSw(x_{n-1-l},y_{q_i-1-l})\\
F_3^{(i)}&:=&\prod_{k=0}^{q_i-q_1-1}\prod_{l=0}^{q_1-1}C_iSw(y_{q_1-l+k},y_{q_1-l-1+k})
\end{eqnarray*}
\end{Definition}

We are now ready to provide the quantum circuit for $f_i$. We gathered the results inside the following theorem.
\begin{Theorem} Quantum circuit for $f_i$ with $2\leq i\leq k$.\\
If $q_i=q_1$, just do nothing. Otherwise, operate as follows:
$$f_i=\begin{cases}
D_0^{(i)}D_1^{(i)}D_2^{(i)}&\text{if $q_i<\frac{q_1}{2}$}\\
D_1^{(i)}D_3^{(i)}D_2^{(i)}&\text{if $q_i=\frac{q_1}{2}$}\\
E_0^{(i)}E_1^{(i)}E_2^{(i)}E_3^{(i)}&\text{if $\frac{q_1}{2}<q_i<q_1$}\\
F_1^{(i)}F_2^{(i)}F_3^{(i)}&\text{if $q_i>q_1$}
\end{cases}$$

\end{Theorem}

The next part addresses an exact count of the quantum gates used to perform the quantum baker map. 

\subsection{Circuit depth}
In what follows, we will use the same notations as before. We consider a quantum baker map with parameter $(q_1,q_2,\dots q_k)$, with the $q_i$ non necessarily distinct, and such that $2^n=\sum_{i=1}^k 2^{q_i}$ and $2^{q_r}|\sum_{j=1}^{r-1}2^{q_j}$ for all integer $r$ such that $2\leq r\leq k$.

We recall from the previous part that there is a quantum circuit $\prod_{i=1}^{k}f_i$ that implements the quantum baker map. 
We will denote by $\mc{N}_i$ the number of quantum gates used to perform $f_i$, where $1\leq i\leq k$. The theorem below provides the $\mc{N}_i$'s.
\begin{Theorem} The quantum baker map can be implemented using $\sum_{i=1}^{k}\mc{N}_i$ quantum gates with:
\begin{eqnarray*}
\mc{N}_1&=&\begin{cases}n+q_1&\text{if}\; q_1\leq\frac{n}{2}\\&\\
(n-q_1)(3q_1-n+2)&\text{if}\; q_1>\frac{n}{2}\end{cases}
\\&&\\&&\\&&\nts\nts\nts\nts\nts\nts\nts\nts\nts\nts\nts\nts\nts\nts\nts\nts\nts\nts\nts\nts\nts\nts\nts\nts\nts\nts\nts\nts\nts\nts\nts\nts\nts\nts\nts\nts\text{and for each $i$ with $2\leq i\leq k$:}\\&&\\
\mc{N}_i&=&\begin{cases}
q_1+q_i&\text{if}\; q_i\leq\frac{q_1}{2}\\&\\
(q_1-q_i)(3q_i-q_1+2)&\text{if}\;\frac{q_1}{2}<q_i\leq q_1\\&\\
(q_i-q_1)(2q_1+1)&\text{if}\;q_i>q_1
\end{cases}.\end{eqnarray*}
\end{Theorem}

\subsection{Analysis and some conclusions}

In this part, we compare on some examples the present scheme and the scheme of \cite{LE}. The comparison centers on the number of qubits needed and the depth of the quantum circuit. The current protocol was introduced in order to reduce the number of qubits by forming blocks of images when the number of images to encrypt exceeds the number of bit planes. However, in some cases of use, while the number of qubits is only slightly decreased, the depth of the quantum circuit can be augmented by a factor $1.8$. In other cases, while the number of qubits is significantly decreased, the depth of the quantum circuit can be augmented by a factor $2.3$. In yet other cases, the numbers are almost identical. We will consider two types of schemes, which we call respectively ``simplified scheme" and ``non-simplified scheme". The non-simplified scheme is the one presented in $\S\, 2.1$ and in \cite{LE}, in the case of a scheme without blocks. 

\textbf{Simplified scheme:} we apply the same QBM at each pixel position, or at each pixel position and block; we apply the same QBM at each image and bit plane, or at each image, bit plane and block. 

\textbf{Non-simplified scheme:} we apply controlled QBM gates for each pixel position, or for each pixel position and each block, with the control depending on the pixel position (resp the pixel position and the block); we apply controlled QBM gates for each bit plane and image, or for each bit plane and image belonging to a certain block, with the control depending on the bit plane and the image or on the bit plane, the image and the block. For each controlled gate, the parameters of the QBM are different. Below, we computed the depth of the quantum circuit corresponding to only one decomposition, that is corresponding to only one set of parameters of the QBM for the square quantum transform involved. Because our goal in the following discussion is simply to get a count estimate using some examples, we will consider that, on average, the number of quantum gates used for each controlled QBM gate is the latter depth. 

In what follows, let $L$ denote the number of bit planes, $M$ denote the number of images to encrypt and $t$ denote the number of blocks. We study the following special cases: $$\begin{array}{cc}(i):&\lb L=8; M=30; t=4\rb,\\(ii):&\lb L=8; M=64; t=8\rb,\\(iii):&\lb L=8; M=128; t=16\rb,\\(iv):&\lb L=8;M=200; t=32\rb.\end{array}$$

We call Protocol I ($P_1$ in the forthcoming equations and table) the protocol of \cite{LE} which does not involve any blocks. We call Protocol II ($P_2$ in the forthcoming equations and table) the protocol of the present paper. 

We recall that in Protocol I, the number of qubits needed is:
$$2(n+\lceil log_2M\rceil)+1,$$
and in Protocol II, the number of qubits needed is:
$$2n+\lceil log_2M\rceil +\lceil log_2L\rceil+1,$$
for an image of size $2^n\times 2^n$, and assuming that the number of images $M$ is larger than the number of bit planes $L$ (usually $8$). In our examples, the images all have the standard size $256\times 256$. The numbers of qubits above are sometimes called the ``width of the quantum circuit". 

In Protocol II, we must apply a QBM on a square of size $8$ and we chose the arbitrary admissible decomposition:
$$2^3=2^2+2+2.$$
With the QBM parameters $q_1=2$ and $q_2=q_3=1$, we compute, using the expressions of Theorem 4:
$$\mathcal{N}_1=5\,\text{and}\,\forall i\in\lb 2,3\rb,\,\N_i=3,$$
yielding a total of $\N=11$ quantum gates. 

In Protocol I, we use a QBM on a square of size $2^{\lc log_2M\rc}$. We choose the arbitrary following admissible decompositions, respectively associated with $(i),\,(ii),\,(iii)$ and $(iv)$:
\begin{eqnarray*} (i): 2^5&=&2^3+2^2+2^2+2^2+2^2+2+2+2+2,\\(ii):2^6&=&2^5+2+2+2^2+2^3+2^4,\\(iii):2^7&=&2^4+2^3+2^3+2^3+2^2+2^2+2^3+2^2+2^2+2^6,\\(iv):2^{8}&=&2^6+2^6+2^2+2^2+2^3+2^4+2^5+2^6.\end{eqnarray*}

And like before, by using Theorem $4$, we compute the total number $\N$ of quantum gates needed to implement the QBM with these respective parameters. We obtain respectively:
$$\N=48,45,91,76\;\text{quantum gates}.$$
The admissible decomposition for $(iv)$ is also the one we use for the action of the QBM on the square of size $256$ containing the pixel positions. We must thus add $76$ to the numbers above in order to obtain the total depth of the quantum circuit in the case of the simplified scheme. In the case of the non-simplified scheme, there are many more quantum gates involved since we use controlled QBM gates instead. In what follows, we define: 
$$\begin{array}{l}n_1:=\text{Average number of gates for shuffling the pixel positions,}\\
n_2:=\text{Average number of gates for shuffling a square of area $2^{2\lc log_2M\rc}$,}\\
n_3:=\text{Average number of gates for shuffling $2^{\lc log_2L\rc}$ bit planes and images}. \end{array}$$
Then, in the non-simplified scheme, the depth of the quantum circuit can be computed as follows, for each of the two protocols, where we still assume that $M>L$ and the images have size $2^n\times 2^n$. 
\begin{eqnarray*}
\# \text{quantum gates}(P_1)&=&n_1. 2^{2\lc log_2M\rc}+n_2. 2^{2n}\\
\# \text{quantum gates}(P_2)&=&n_1. 2^{\lM+\lL}+n_3.2^{2n}\times 2^{\lM-\lL}
\end{eqnarray*}

The numerical results in terms of number of qubits used and number of quantum gates needed are displayed in Table $1$ below, for both the simplified scheme and the non-simplified one, in the special cases studied here. 
\begin{center}
\begin{tabular}{cc|c|c|c|c|}
&&$M=30$&$M=64$&$M=128$&$M=200$\\\hline
$\nts\nts\nts\nts\nts\#$ qubits&$\nts\nts\nts\nts\nts\begin{array}{l}P_1\\P_2\end{array}$&$\begin{array}{l}27\\25\end{array}$&$\begin{array}{l}29\\26\end{array}$&$\begin{array}{l}31\\27\end{array}$&$\begin{array}{l}33\\28\end{array}$\\\hline
$\begin{array}{l}\nts\nts\nts\nts\nts\textit{Simplified}\\\nts\nts\nts\#\text{gates}\end{array}$&$\nts\nts\nts\nts\nts\begin{array}{l}P_1\\P_2\end{array}$&$\begin{array}{l}124\\\;87\end{array}$&$\begin{array}{l}121\\\;87\end{array}$&$\begin{array}{l}167\\\;87\end{array}$&$\begin{array}{l}152\\\;87\end{array}$\\\hline
$\begin{array}{l}\nts\nts\nts\nts\nts\textit{Non-simplified}\\\#\text{gates}\end{array}$&$\nts\nts\nts\nts\nts\begin{array}{l}P_1\\P_2\end{array}$&$\begin{array}{l}3\,223\,552\\2\,903\,040\end{array}$&$\begin{array}{l}3\,260\,416\\5\,806\,080\end{array}$&$\begin{array}{l}7\,208\,960\\11\,612\,160\end{array}$&$\begin{array}{l}9\,961\,472\\23\,224\,320\end{array}$\\\hline $\begin{array}{l}\nts\nts\nts\textit{Non-simplified}\\\#(P_2)/\#(P_1)\end{array}$&&0.9&1.7&1.6&2.3\\\hline
\end{tabular}
\end{center}
\hfill \textit{Table $1$.}\\
In the naive version in which no controlled QBM gates are used, the scheme by blocks only presents advantages, namely it reduces the number of qubits needed and at the same time uses fewer quantum gates. However, it seems of course much less secure than realizing an independent scrambling of the pixel positions (resp of the bit plane and the images) for each bit plane and image, or bit plane and image belonging to a certain block (resp for each pixel position, or each pixel position and each block), using controlled QBM gates, although the precise data issued from a cryptanalysis are not yet available. 

By reading Table $1$, we see that for reducing significantly the width of the quantum circuit, we must increase its depth for the most secure version presented in this paper. It appears worth to use the protocol with blocks of this paper only when the number of images to encrypt is quite large. Then, work still needs to be achieved in order to optimize the quantum circuit depth without harming the security of the cryptographic scheme. There exist many possibilities for achieving such a good balance. 

Last, we raise the question whether we could achieve in an intermediate type scheme (lying in between the simplified scheme and the non-simplified one, both discussed here) a realizable quantum circuit with lower depth than the corresponding quantum circuit of \cite{LE} and leading to good security performances for encrypting a large number of images using blocks. \\


\newpage

\begin{Large}Appendix. \end{Large}The sine chaotification of the 5D hyperchaotic system. \\\\
With parameters set to $(\la_1,\la_2,\la_3,\la_4,\la_5)=(49,23,58,120,237)$ and initial conditions set to $(0.1,0.5,0.2,-0.8,0.9)$, we obtain the following plot after $250$ iterations. 
$$\begin{array}{l}\\
\end{array}$$

\begin{center}
\epsfig{file=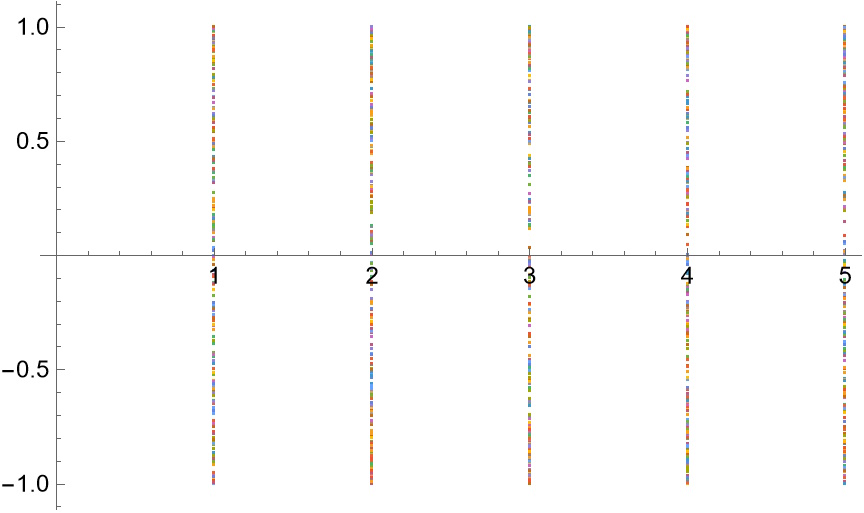, height=7cm}
\end{center}


\newpage

\begin{thebibliography}{ll}
\bibitem{BA} C. Hou and X. Liu, Quantum image scrambling algorithm based on discrete Baker map, Modern Physics Letters A $(2020)$
\bibitem{HZ} Z. Hua, B. Zhou, Y. Zhou, Sine chaotification model for enhancing chaos and its hardware implementation, IEE Transactions on Industrial Electronics, Vol. $66$, Issue $2$ 
$(2019)$ $1273-1284$
\bibitem{LE} C. Levaillant, Quantum multiple gray scale images encryption scheme in the bit plane representation model, arXiv:$2401.00787$
\bibitem{BP} X. Liu and C. Liu, Quantum image encryption scheme using independent bit-plane permutation and Baker map, Quantum Information Processing Vol. $22$ $262$ $(2023)$
\bibitem{QB} X. Liu, D. Xiao, W. Huang and C. Liu, Quantum block image encryption based on Arnold transform and sine chaotification model, IEE Access Vol. 7 $(2019)$ $55188-55199$
\bibitem{MA} R. Matthews, On the derivation of a "chaotic" encryption algorithm, Cryptologia $(1989)$ Vol. 13 Issue 1 $29-42$
\bibitem{SS} S. Sun, A novel hyperchaotic image encryption scheme based on DNA encoding, pixel-level scrambling and bit-level scrambling, IEE Photonics Journal, Vol. $10$, No.$2$ $(2018)$
\bibitem{YZ} Q. Yang, D. Zhu, and L. Yang, A new 7D hyperchaotic system with five positive Lyapunov exponents coined, Int. J. Bifurcation Chaos, Vol. 28, no. 5 (2018), art. no. 1850057
\bibitem{NE} Y. Zhang, K. Lu, Y.H. Gao, M. Wang, NEQR: a novel enhanced quantum representation of digital images, Quantum Information Processing Vol. 12 $(2013)$ $2833-2860$
\bibitem{ZH} N. Zhou, X. Yan, H. Liang, X. Tao and G. Li, Multi-image encryption scheme based on quantum $3$D Arnold transform and scaled Zhongtang chaotic system, Quantum Information Processing $(2018)$ $17:338$
\end{thebibliography}
\end{document}